# A short introduction to local fractional complex analysis

Yang Xiao-Jun

*Department of Mathematics and Mechanics, China University of Mining and Technology, Xuzhou Campus, Xuzhou, Jiangsu, 221008, P. R.C*

dyangxiaojun@163.com

This paper presents a short introduction to local fractional complex analysis. The generalized local fractional complex integral formulas, Yang-Taylor series and local fractional Laurent's series of complex functions in complex fractal space, and generalized residue theorems are investigated.

## 1 Introduction

Local fractional calculus has played an important role in not only mathematics but also in physics and engineers [1-12]. There are many definitions of local fractional derivatives and local fractional integrals (also called fractal calculus). Hereby we write down local fractional derivative, given by [5-7]

$$f^{(\alpha)}(x_0) = \frac{d^\alpha f(x)}{dx^\alpha}\bigg|_{x=x_0} = \lim_{x \to x_0} \frac{\Delta^\alpha(f(x) - f(x_0))}{(x-x_0)^\alpha} \quad (1.1)$$

with $\Delta^\alpha(f(x) - f(x_0)) \cong \Gamma(1+\alpha)\Delta(f(x) - f(x_0))$, and local fractional integral of $f(x)$, denoted by [5-6,8]

$$_aI_b^{(\alpha)}f(x) = \frac{1}{\Gamma(1+\alpha)}\int_a^b f(t)(dt)^\alpha = \frac{1}{\Gamma(1+\alpha)}\lim_{\Delta t \to 0}\sum_{j=0}^{j=N-1} f(t_j)(\Delta t_j)^\alpha \quad (1.2)$$

with $\Delta t_j = t_{j+1} - t_j$ and $\Delta t = \max\{\Delta t_1, \Delta t_2, \Delta t_j, ...\}$, where for $j = 0, ..., N-1$, $[t_j, t_{j+1}]$ is a partition of the interval $[a,b]$ and $t_0 = a, t_N = b$.

More recently, a motivation of local fractional derivative and local fractional integral of complex functions is given [11]. Our attempt, in the present paper, is to continue to study local fractional calculus of complex function. As well, a short outline of local fractional complex analysis will be established.

# 2 Local fractional calculus of the complex-variable functions

In this section we deduce fundamentals of local fractional calculus of the complex-valued functions. Here we start with local fractional continuity of complex functions.

## 2.1 Local fractional continuity of complex-variable functions

**Definition 1**
Given $z_0$ and $|z - z_0| < \delta$, then for any $z$ we have [11]

$$|f(z) - f(z_0)| < \varepsilon^\alpha. \tag{2.1}$$

Here complex function $f(z)$ is called local fractional continuous at $z = z_0$, denoted by

$$\lim_{z \to z_0} f(z) = f(z_0). \tag{2.2}$$

A function $f(z)$ is called local fractional continuous on the region $\Re$, denoted by

$$f(z) \in C_\alpha(\Re).$$

As a direct result, we have the following results:

Suppose that $\lim_{z \to z_0} f(z) = f(z_0)$ and $\lim_{z \to z_0} g(z) = g(z_0)$, then we have that

$$\lim_{z \to z_0} [f(z) \pm g(z)] = f(z_0) \pm g(z_0), \tag{2.3}$$

$$\lim_{z \to z_0} [f(z) g(z)] = f(z_0) g(z_0), \tag{2.4}$$

and

$$\lim_{z \to z_0} [f(z) / g(z)] = f(z_0) / g(z_0), \tag{2.5}$$

the last only if $g(z_0) \neq 0$.

## 2.2 Local fractional derivatives of complex function

**Definition 2**
Let the complex function $f(z)$ be defined in a neighborhood of a point $z_0$. The local fractional derivative of $f(z)$ at $z_0$ is defined by the expression [11]

$$_{z_0}D_z^\alpha f(z) =: \lim_{z \to z_0} \frac{\Gamma(1+\alpha)[f(z) - f(z_0)]}{(z - z_0)^\alpha}, 0 < \alpha \leq 1. \tag{2.6}$$

If this limit exists, then the function $f(z)$ is called to be local fractional analytic at $z_0$, denoted by

$$_{z_0}D_z^\alpha f(z), \left.\frac{d^\alpha}{dz^\alpha} f(z)\right|_{z=z_0} \text{ or } f^{(\alpha)}(z_0).$$

**Remark 1.** If the limits exist for all $z_0$ in a region $\Re$, then $f(z)$ is said to be local fractional analytic in a region $\Re$, denoted by

$$f(z) \in D(\Re).$$

Suppose that $f(z)$ and $g(z)$ are local fractional analytic functions, the following rules are valid [11].

$$\frac{d^\alpha (f(z) \pm g(z))}{dz^\alpha} = \frac{d^\alpha f(z)}{dz^\alpha} \pm \frac{d^\alpha g(z)}{dz^\alpha}; \tag{2.7}$$

$$\frac{d^\alpha (f(z) g(z))}{dz^\alpha} = g(z)\frac{d^\alpha f(z)}{dz^\alpha} + f(z)\frac{d^\alpha g(z)}{dz^\alpha}; \tag{2.8}$$

$$\frac{d^\alpha \left(\frac{f(z)}{g(z)}\right)}{dz^\alpha} = \frac{g(z)\frac{d^\alpha f(z)}{dz^\alpha} + f(z)\frac{d^\alpha g(z)}{dz^\alpha}}{g(z)^2} \tag{2.9}$$

if $g(z) \neq 0$;

$$\frac{d^\alpha (Cf(z))}{dz^\alpha} = C\frac{d^\alpha f(z)}{dz^\alpha}, \tag{2.10}$$

where $C$ is a constant;

If $y(z) = (f \circ u)(z)$ where $u(z) = g(z)$, then

$$\frac{d^\alpha y(z)}{dz^\alpha} = f^{(\alpha)}(g(z))(g^{(1)}(z))^\alpha. \tag{2.11}$$

## 2.3 Local fractional Cauchy-Riemann equations

**Definition 3**
If there exists a function

$$f(z) = u(x, y) + i^\alpha v(x, y), \tag{2.12}$$

where $u$ and $v$ are real functions of $x$ and $y$. The local fractional complex differential equations

$$\frac{\partial^\alpha u(x, y)}{\partial x^\alpha} - \frac{\partial^\alpha v(x, y)}{\partial y^\alpha} = 0 \tag{2.13}$$

and

$$\frac{\partial^\alpha u(x, y)}{\partial y^\alpha} + \frac{\partial^\alpha v(x, y)}{\partial x^\alpha} = 0 \tag{2.14}$$

are called local fractional Cauchy-Riemann Equations.

**Theorem 1**
Suppose that the function

$$f(z) = u(x, y) + i^\alpha v(x, y) \tag{2.15}$$

is local fractional analytic in a region $\Re$. Then we have

$$\frac{\partial^\alpha u(x,y)}{\partial x^\alpha} - \frac{\partial^\alpha v(x,y)}{\partial y^\alpha} = 0 \qquad (2.16)$$

and

$$\frac{\partial^\alpha u(x,y)}{\partial y^\alpha} + \frac{\partial^\alpha v(x,y)}{\partial x^\alpha} = 0. \qquad (2.17)$$

*Proof.* Since $f(z) = u(x,y) + i^\alpha v(x,y)$, we have the following identity

$$f^{(\alpha)}(z_0) = \lim_{z \to z_0} \frac{\Gamma(1+\alpha)\left[f(z) - f(z_0)\right]}{(z - z_0)^\alpha}. \qquad (2.18)$$

Consequently, the formula (2.18) implies that

$$\lim_{\Delta z \to 0} \frac{\Gamma(1+\alpha)\left[f(z+\Delta z) - f(z)\right]}{\Delta z^\alpha}$$
$$= \lim_{\substack{\Delta x \to 0 \\ \Delta y \to 0}} \frac{\Gamma(1+\alpha)\left[u(x+\Delta x, y+\Delta y) - u(x,y) + i^\alpha \left(v(x+\Delta x, y+\Delta y) - v(x,y)\right)\right]}{\Delta x^\alpha + i^\alpha \Delta y^\alpha}. \qquad (2.19)$$

In a similar manner, setting $\Delta y \to 0$ and taking into account the formula (2.19), we have $(\Delta y)^\alpha \to 0$ such that

$$f^{(\alpha)}(z_0) = \lim_{\Delta y \to 0} \frac{\Gamma(1+\alpha)\left[u(x, y+\Delta y) - u(x,y) + i^\alpha \left(v(x, y+\Delta y) - v(x,y)\right)\right]}{i^\alpha \Delta y^\alpha}. \qquad (2.20)$$

Hence

$$f^{(\alpha)}(z_0) = -i^\alpha \frac{\partial^\alpha u(x,y)}{\partial y^\alpha} + \frac{\partial^\alpha v(x,y)}{\partial y^\alpha} \qquad (2.21)$$

If $\Delta x \to 0$, from (2.19) we have $(\Delta x)^\alpha \to 0$ such that

$$f^{(\alpha)}(z_0) = \lim_{\Delta x \to 0} \frac{\Gamma(1+\alpha)\left[u(x+\Delta x, y) - u(x,y) + i^\alpha \left(v(x+\Delta x, y) - v(x,y)\right)\right]}{\Delta x^\alpha} \qquad (2.22)$$

Thus we get the identity

$$f^{(\alpha)}(z_0) = \frac{\partial^\alpha u(x,y)}{\partial x^\alpha} + i^\alpha \frac{\partial^\alpha v(x,y)}{\partial x^\alpha}. \qquad (2.24)$$

Since $f(z) = u(x,y) + i^\alpha v(x,y)$ is local fractional analytic in a region $\Re$, we have the following formula

$$f^{(\alpha)}(z_0) = \frac{\partial^\alpha u(x,y)}{\partial x^\alpha} + i^\alpha \frac{\partial^\alpha v(x,y)}{\partial x^\alpha} = -i^\alpha \frac{\partial^\alpha u(x,y)}{\partial y^\alpha} + \frac{\partial^\alpha v(x,y)}{\partial y^\alpha}. \qquad (2.25)$$

Hence, from (2.25), we arrive at the following identity

$$\frac{\partial^\alpha u(x,y)}{\partial x^\alpha} - \frac{\partial^\alpha v(x,y)}{\partial y^\alpha} = 0 \qquad (2.26)$$

and

$$\frac{\partial^\alpha u(x,y)}{\partial y^\alpha} + \frac{\partial^\alpha v(x,y)}{\partial x^\alpha} = 0. \qquad (2.27)$$

This completes the proof of Theorem 1.

**Remark 2.** Local fractional C-R equations are sufficient conditions that $f(z)$ is local fractional analytic in $\Re$.

The local fractional partial equations

$$\frac{\partial^{2\alpha} u(x,y)}{\partial x^{2\alpha}} + \frac{\partial^{2\alpha} u(x,y)}{\partial y^{2\alpha}} = 0 \qquad (2.28)$$

and

$$\frac{\partial^{2\alpha} v(x,y)}{\partial x^{2\alpha}} + \frac{\partial^{2\alpha} v(x,y)}{\partial y^{2\alpha}} = 0 \qquad (2.29)$$

are called local fractional Laplace equations, denoted by

$$\nabla^\alpha u(x,y) = 0 \qquad (2.30)$$

and

$$\nabla^\alpha v(x,y) = 0, \qquad (2.31)$$

where

$$\nabla^\alpha = \frac{\partial^{2\alpha}}{\partial x^{2\alpha}} + \frac{\partial^{2\alpha}}{\partial y^{2\alpha}} \qquad (2.32)$$

is called local fractional Laplace operator.

**Remark 3.** Suppose that $\nabla^\alpha u(x,y) = 0$, $u(x,y)$ is a local fractional harmonic function in $\Re$.

## 2.4 Local fractional integrals of complex function

**Definition 4**
Let $f(z)$ be defined, single-valued and local fractional continuous in a region $\Re$. The local fractional integral of $f(z)$ along the contour $C$ in $\Re$ from point $z_p$ to point $z_q$, is defined as [11]

$$\begin{aligned} I_C^\alpha f(z) &= \frac{1}{\Gamma(1+\alpha)} \lim_{\Delta z \to 0} \sum_{i=0}^{n-1} f(z_i)(\Delta z)^\alpha \\ &= \frac{1}{\Gamma(1+\alpha)} \int_C f(z)(dz)^\alpha \end{aligned} \qquad (2.33)$$

where for $i = 0, 1, \ldots, n$ $\Delta z_i = z_i - z_{i-1}$, $z_0 = z_p$ and $z_n = z_q$.

For convenience, we assume that

$$_{z_0}I_{z_0}^{(\alpha)} f(z) = 0 \tag{2.34}$$

if $z = z_0$.

The rules for complex integration are similar to those for real integrals. Some important results are as follows [11]:

Suppose that $f(z)$ and $g(z)$ be local fractional continuous along the contour $C$ in $\Re$.

$$\frac{1}{\Gamma(1+\alpha)}\int_C (f(z)+g(z))(dz)^\alpha = \frac{1}{\Gamma(1+\alpha)}\int_C f(z)(dz)^\alpha + \frac{1}{\Gamma(1+\alpha)}\int_C g(z)(dz)^\alpha; \tag{2.35}$$

$$\frac{1}{\Gamma(1+\alpha)}\int_C kf(z)(dz)^\alpha = \frac{k}{\Gamma(1+\alpha)}\int_C f(z)(dz)^\alpha, \tag{2.36}$$

for a constant $k$;

$$\frac{1}{\Gamma(1+\alpha)}\int_C f(z)(dz)^\alpha = \frac{1}{\Gamma(1+\alpha)}\int_{C_1} f(z)(dz)^\alpha + \frac{1}{\Gamma(1+\alpha)}\int_{C_2} f(z)(dz)^\alpha, \tag{2.37}$$

where $C = C_1 + C_2$;

$$\frac{1}{\Gamma(1+\alpha)}\int_{C_1} f(z)(dz)^\alpha = -\frac{1}{\Gamma(1+\alpha)}\int_{-C_1} f(z)(dz)^\alpha; \tag{2.38}$$

$$\left|\frac{1}{\Gamma(1+\alpha)}\int_C f(z)(dz)^\alpha\right| \leq \frac{1}{\Gamma(1+\alpha)}\int_C |f(z)|\left|(dz)^\alpha\right| \leq ML, \tag{2.39}$$

where $M$ is an upper bound of $f(z)$ on $C$ and $L = \frac{1}{\Gamma(1+\alpha)}\int_C \left|(dz)^\alpha\right|$.

**Theorem 2**

If the contour $C$ has end points $z_p$ and $z_q$ with orientation $z_p$ to $z_q$, and if function $f(z)$ has the primitive $F(z)$ on $C$, then we have

$$\frac{1}{\Gamma(1+\alpha)}\int_C f(z)(dz)^\alpha = F(z_q) - F(z_p). \tag{2.40}$$

***Remark 4.*** *Suppose that* $f(z) \in D(\Re)$. *For* $k = 0, 1, \ldots, n$ *and* $0 < \alpha \leq 1$ *there exists a local fractional series*

$$f(z) = \sum_{k=0}^{\infty} \frac{f^{(k\alpha)}(z_0)}{\Gamma(1+k\alpha)}(z-z_0)^{k\alpha} \tag{2.41}$$

with $f^{(k\alpha)}(z) \in D(\Re)$, where $f^{(k\alpha)}(z) = \overbrace{D_z^{(\alpha)} \ldots D_z^{(\alpha)}}^{k\ times} f(z)$.

This series is called Yang-Taylor series of local fractional analytic function (for real function case, see [12].)

**Theorem 3**

If $C$ is a simple closed contour, and if function $f(z)$ has a primitive on $C$, then [11]

$$\frac{1}{\Gamma(1+\alpha)}\oint_C f(z)(dz)^\alpha = 0. \tag{2.42}$$

**Corollary 4**

If the closed contours $C_1$, $C_2$ is such that $C_2$ lies inside $C_1$, and if $f(z)$ is local fractional analytic on $C_1$, $C_2$ and between them, then we have [11]

$$\frac{1}{\Gamma(1+\alpha)}\int_{C_1} f(z)(dz)^\alpha = \frac{1}{\Gamma(1+\alpha)}\int_{C_2} f(z)(dz)^\alpha. \tag{4.43}$$

**Theorem 5**

Suppose that the closed contours $C_1, C_2$ is such that $C_2$ lies inside $C_1$, and if $f(z)$ is local fractional analytic on $C_1, C_2$ and between them, then we have[11]

$$\frac{1}{\Gamma(1+\alpha)}\int_{C_1} f(z)(dz)^\alpha = \frac{1}{\Gamma(1+\alpha)}\int_{C_2} f(z)(dz)^\alpha. \tag{2.44}$$

# 3 Generalized local fractional integral formulas of complex functions

In this section we start with generalized local fractional integral formulas of complex functions and deduce some useful results.

**Theorem 6**

Suppose that $f(z)$ is local fractional analytic within and on a simple closed contour $C$ and $z_0$ is any point interior to $C$. Then we have

$$\frac{1}{(2\pi)^\alpha i^\alpha}\cdot\frac{1}{\Gamma(1+\alpha)}\oint_C \frac{f(z)}{(z-z_0)^\alpha}(dz)^\alpha = f(z_0). \tag{3.1}$$

*Proof.* From (2.44), we arrive at the formula

$$\frac{1}{(2\pi)^\alpha i^\alpha}\cdot\frac{1}{\Gamma(1+\alpha)}\oint_C \frac{f(z)}{(z-z_0)^\alpha}(dz)^\alpha = \frac{1}{(2\pi)^\alpha i^\alpha}\cdot\frac{1}{\Gamma(1+\alpha)}\oint_{C_1} \frac{f(z)}{(z-z_0)^\alpha}(dz)^\alpha, \tag{3.2}$$

where $C_1 : \left|(z-z_0)^\alpha\right| = \varepsilon^\alpha$.

Setting $\left|(z-z_0)^\alpha\right| = \varepsilon^\alpha$ implies that

$$z^\alpha - z_0^\alpha = \varepsilon^\alpha E_\alpha(i^\alpha \theta^\alpha) \tag{3.3}$$

and
$$(dz)^\alpha = i^\alpha \varepsilon^\alpha E_\alpha\left(i^\alpha \theta^\alpha\right)(d\theta)^\alpha. \tag{3.4}$$

Taking (3.3) and (3.4), it follows from (3.2) that

$$\frac{1}{(2\pi)^\alpha i^\alpha} \cdot \frac{1}{\Gamma(1+\alpha)} \int_0^{2\pi} \frac{f\left(z_0 + \varepsilon E(i\theta)\right)}{\varepsilon^\alpha E_\alpha\left(i^\alpha \theta^\alpha\right)} i^\alpha \varepsilon^\alpha E_\alpha\left(i^\alpha \theta^\alpha\right)(d\theta)^\alpha$$
$$= \lim_{\varepsilon \to 0} \frac{1}{(2\pi)^\alpha} \cdot \frac{1}{\Gamma(1+\alpha)} \int_0^{2\pi} f\left(z_0 + \varepsilon E(i\theta)\right)(d\theta)^\alpha \tag{3.5}$$

From (3.5), we get

$$\frac{1}{(2\pi)^\alpha} \cdot \frac{1}{\Gamma(1+\alpha)} \int_0^{2\pi} \left(\lim_{\varepsilon \to 0} f\left(z_0 + \varepsilon E(i\theta)\right)\right)(d\theta)^\alpha = \frac{f(z_0)}{(2\pi)^\alpha} \cdot \frac{1}{\Gamma(1+\alpha)} \int_0^{2\pi} (d\theta)^\alpha \tag{3.6}$$

Furthermore

$$\frac{f(z_0)}{(2\pi)^\alpha} \cdot \frac{1}{\Gamma(1+\alpha)} \int_0^{2\pi} (d\theta)^\alpha = f(z_0). \tag{3.7}$$

Substituting (3.7) into (3.6) and (3.3) implies that

$$\frac{1}{(2\pi)^\alpha i^\alpha} \cdot \frac{1}{\Gamma(1+\alpha)} \oint_C \frac{f(z)}{(z-z_0)^\alpha} (dz)^\alpha = f(z_0). \tag{3.8}$$

The proof of the theorem is completed.

Likewise, we have the following corollary:

**Corollary 7**

Suppose that $f(z)$ is local fractional analytic within and on a simple closed contour $C$ and $z_0$ is any point interior to $C$. Then we have

$$\frac{1}{(2\pi)^\alpha i^\alpha} \cdot \frac{1}{\Gamma(1+\alpha)} \oint_C \frac{f(z)}{(z-z_0)^{(n+1)\alpha}} (dz)^\alpha = f^{(n\alpha)}(z_0). \tag{3.9}$$

*Proof.* Taking into account formula (3.1), we arrive at the identity.

**Theorem 8**

Suppose that $f(z)$ is local fractional analytic within and on a simple closed contour $C$ and $z_0$ is any point interior to $C$. Then we have

$$\frac{1}{\Gamma(1+\alpha)} \oint_C \frac{(dz)^\alpha}{(z-z_0)^\alpha} = (2\pi)^\alpha i^\alpha. \tag{3.9}$$

*Proof.* Taking $f(z) = 1$, from (3.9) we deduce the result.

**Theorem 9**

Suppose that $f(z)$ is local fractional analytic within and on a simple closed contour $C$ and $z_0$ is any point interior to $C$. Then we have

$$\frac{1}{\Gamma(1+\alpha)}\oint_C \frac{(dz)^\alpha}{(z-z_0)^{n\alpha}}=0, \text{ for } n>1. \tag{3.10}$$

*Proof.* Taking $f(z)=1$, from (3.9) we deduce the result.

# 4 Complex Yang-Taylor's series and local fractional Laurent's series

In this section we start with a Yang-Taylor's expansion formula of complex functions and deduce local fractional Laurent series of complex functions.

## 4.1 Complex Yang-Taylor's expansion formula

**Definition 5**
Let $f(z)$ be local fractional analytic inside and on a simple closed contour $C$ having its center at $z=z_0$. Then for all points $z$ in the circle we have the Yang-Taylor series representation of $f(z)$, given by

$$\begin{aligned}f(z)&=f(z_0)+\frac{f^{(\alpha)}(z_0)}{\Gamma(1+\alpha)}(z-z_0)^\alpha+\\&\frac{f^{(2\alpha)}(z_0)}{\Gamma(1+2\alpha)}(z-z_0)^{2\alpha}+....+\frac{f^{(k\alpha)}(z_0)}{\Gamma(1+k\alpha)}(z-z_0)^{k\alpha}+...\end{aligned} \tag{4.1}$$

For $C:|z-z_0|^\alpha \leq R^\alpha$, we have the complex Yang-Taylor series

$$f(z)=\sum_{k=0}^{\infty}a_k(z-z_0)^{k\alpha}. \tag{4.2}$$

From (3.44) the above expression implies

$$a_k=\frac{1}{(2\pi)^\alpha i^\alpha}\cdot\frac{1}{\Gamma(1+\alpha)}\oint_C\frac{f(z)}{(z-z_0)^{(k+1)\alpha}}(dz)^\alpha=\frac{f^{(k\alpha)}(z_0)}{\Gamma(1+k\alpha)}, \tag{4.3}$$

for $c:|z-z_0|^\alpha \leq R^\alpha$.

Successively, it follows from (4.3) that

$$f(z)=\sum_{k=0}^{\infty}a_k(z-z_0)^{k\alpha}, \tag{4.4}$$

where

$$a_k=\frac{1}{(2\pi)^\alpha i^\alpha}\cdot\frac{1}{\Gamma(1+\alpha)}\oint_C\frac{f(z)}{(z-z_0)^{(k+1)\alpha}}(dz)^\alpha=\frac{f^{(k\alpha)}(z_0)}{\Gamma(1+k\alpha)}, \tag{4.5}$$

for $C : |z - z_0|^\alpha \leq R^\alpha$.

Hence, the above formula implies the relation (4.2).

**Theorem 10**

Suppose that complex function $f(z)$ is local fractional analytic inside and on a simple closed contour $C$ having its center at $z = z_0$. There exist all points $z$ in the circle such that we have the Yang-Taylor's series of $f(z)$

$$f(z) = \sum_{k=0}^{\infty} a_k (z - z_0)^{k\alpha}, \qquad (4.5)$$

where

$$a_k = \frac{1}{(2\pi)^\alpha i^\alpha} \cdot \frac{1}{\Gamma(1+\alpha)} \oint_C \frac{f(z)}{(z-z_0)^{(k+1)\alpha}} (dz)^\alpha = \frac{f^{(k\alpha)}(z_0)}{\Gamma(1+k\alpha)},$$

for $C : |z - z_0|^\alpha \leq R^\alpha$.

*Proof.* Setting $C_1 : |z - z_0|^\alpha = R^\alpha$ and using (3.1), we have

$$f(z) = \frac{1}{(2\pi)^\alpha i^\alpha} \cdot \frac{1}{\Gamma(1+\alpha)} \oint_{C_1} \frac{f(\xi)}{(\xi-z)^\alpha} (d\xi)^\alpha. \qquad (4.6)$$

Taking $\xi \in C_1$, we get

$$\frac{|z - z_0|^\alpha}{|\xi - z_0|^\alpha} = q^\alpha < 1 \qquad (4.7)$$

and

$$\begin{aligned}
\frac{1}{(\xi - z)^\alpha} &= \frac{1}{(\xi - z_0)^\alpha} \frac{1}{1 - \frac{(z - z_0)^\alpha}{(\xi - z_0)^\alpha}} \\
&= \frac{1}{(\xi - z_0)^\alpha} \frac{1}{1 - \left(\frac{z - z_0}{\xi - z_0}\right)^\alpha} \\
&= \sum_{n=1}^{\infty} \frac{1}{(\xi - z_0)^{(n+1)\alpha}} (z - z_0)^{n\alpha}.
\end{aligned} \qquad (4.8)$$

Substituting (4.8) into (4.6) implies that

$$f(z)$$
$$= \sum_{n=1}^{N} \left[ \frac{1}{(2\pi)^{\alpha} i^{\alpha}} \cdot \frac{1}{\Gamma(1+\alpha)} \oint_{C_1} \frac{f(\xi)(d\xi)^{\alpha}}{(\xi - z_0)^{(n+1)\alpha}} \right] (z - z_0)^{n\alpha} \quad (4.9)$$
$$+ \frac{1}{(2\pi)^{\alpha} i^{\alpha}} \cdot \frac{1}{\Gamma(1+\alpha)} \oint_{C_1} \sum_{n=N}^{\infty} \left[ \frac{f(\xi)(z - z_0)^{n\alpha}}{(\xi - z_0)^{(n+1)\alpha}} \right] (d\xi)^{\alpha}.$$

Taking the Yang-Taylor formula of analytic function into account, we have the following relation

$$f(z) = \sum_{n=0}^{N-1} \frac{f^{(n\alpha)}(z_0)(z - z_0)^{n\alpha}}{\Gamma(1+n\alpha)} + R_N, \quad (4.10)$$

where $R_N$ is reminder in the form

$$R_N = \frac{1}{(2\pi)^{\alpha} i^{\alpha}} \cdot \frac{1}{\Gamma(1+\alpha)} \oint_{C_1} \sum_{n=N}^{\infty} \left[ \frac{f(\xi)(z - z_0)^{n\alpha}}{(\xi - z_0)^{(n+1)\alpha}} \right] (d\xi)^{\alpha}. \quad (4.11)$$

There exists a Yang-Taylor series

$$f(z) = \sum_{n=0}^{\infty} \frac{f^{(n\alpha)}(z_0)(z - z_0)^{n\alpha}}{\Gamma(1+n\alpha)} \quad (4.12)$$

where is $f(z_0)$ is local fractional analytic at $z = z_0$.

Taking into account the relation $\left| \frac{(z - z_0)^{n\alpha}}{(\xi - z_0)^{n\alpha}} \right| = q^{n\alpha} < 1$ and $|f(z)| \leq M$, from (4.11) we get

$$\begin{aligned}
|R_N| &= \left| \frac{1}{(2\pi)^{\alpha} i^{\alpha}} \cdot \frac{1}{\Gamma(1+\alpha)} \oint_{C_1} \sum_{n=N}^{\infty} \left[ \frac{f(\xi)(z - z_0)^{n\alpha}}{(\xi - z_0)^{(n+1)\alpha}} \right] (d\xi)^{\alpha} \right| \\
&\leq \frac{1}{(2\pi)^{\alpha}} \cdot \frac{1}{\Gamma(1+\alpha)} \oint_{C_1} \left| \sum_{n=N}^{\infty} \frac{|f(\xi)||(z - z_0)^{n\alpha}|}{|(\xi - z_0)^{(n+1)\alpha}|} \right| (d\xi)^{\alpha} \\
&\leq \frac{1}{(2\pi)^{\alpha}} \cdot \frac{1}{\Gamma(1+\alpha)} \oint_{C_1} \left| \sum_{n=N}^{\infty} \frac{|M|}{|(\xi - z_0)^{\alpha}|} \frac{|(z - z_0)^{n\alpha}|}{|(\xi - z_0)^{n\alpha}|} \right| (d\xi)^{\alpha} \\
&\leq \frac{(2\pi)^{\alpha} R^{\alpha}}{(2\pi)^{\alpha}} \cdot \frac{|M|}{\Gamma(1+\alpha)} \frac{q^{n\alpha}}{1 - q^{\alpha}} \\
&\leq \frac{|M| R^{\alpha}}{\Gamma(1+\alpha)} \frac{q^{n\alpha}}{1 - q^{\alpha}}
\end{aligned} \quad (4.13)$$

Furthermore

$$\lim_{N \to \infty} R_N = 0.$$

From (4.9), we have

$$f(z) = \sum_{n=1}^{\infty}\left[\frac{1}{(2\pi)^{\alpha}i^{\alpha}} \cdot \frac{1}{\Gamma(1+\alpha)}\oint_{C_1}\frac{f(\xi)(d\xi)^{\alpha}}{(\xi-z_0)^{(n+1)\alpha}}\right](z-z_0)^{n\alpha}. \qquad (4.14)$$

Hence

$$a_n = \frac{1}{(2\pi)^{\alpha}i^{\alpha}} \cdot \frac{1}{\Gamma(1+\alpha)}\oint_{C_1}\frac{f(\xi)(d\xi)^{\alpha}}{(\xi-z_0)^{(n+1)\alpha}}. \qquad (4.15)$$

Hence the proof of the theorem is completed.

## 4.2 Singular point and poles

**Definition 6**

A singular point of a function $f(z)$ is a value of $z$ at which $f(z)$ fails to be local fractional analytic. If $f(z)$ is local fractional analytic everywhere in some region except at an interior point $z = z_0$, we call $f(z)$ an isolated singularity.

If

$$f(z) = \frac{\phi(z)}{(z-z_0)^{n\alpha}} \qquad (4.16)$$

and

$$\phi(z) \neq 0 \qquad (4.17)$$

where $\phi(z)$ is local fractional analytic everywhere in a region including $z = z_0$, and if $n$ is a positive integer, then $f(z)$ has an isolated singularity at $z = z_0$, which is called a pole of order $n$.

If $n = 1$, the pole is often called a simple pole;

if $n = 2$, it is called a double pole, and so on.

## 4.3 Local fractional Laurent's series

**Definition 7**

If $f(z)$ has a pole of order $n$ at $z = z_0$ but is local fractional analytic at every other point inside and on a contour $C$ with center at $z_0$, then

$$\phi(z) = (z-z_0)^{n\alpha} f(z) \qquad (4.18)$$

is local fractional analytic at all points inside and on $C$ and has a Yang-Taylor series about $z = z_0$ so that

$$f(z)$$
$$= \frac{a_{-n}}{(z-z_0)^{n\alpha}} + \frac{a_{-n+1}}{(z-z_0)^{(n-1)\alpha}} + ... + \quad (4.19)$$
$$\frac{a_{-1}}{(z-z_0)^{\alpha}} + a_0 + a_1(z-z_0)^{\alpha} + .... + a_n(z-z_0)^{n\alpha} + ...$$

This is called a local fractional Laurent series for $f(z)$.

More generally, it follows that

$$f(z) = \sum_{i=-\infty}^{\infty} a_k (z-z_0)^{k\alpha} \quad (4.20)$$

as a local fractional Laurent series.

For $C: r^{\alpha} < |z-z_0|^{\alpha} < R^{\alpha}$ we have a local fractional Laurent series

$$f(z) = \sum_{k=-\infty}^{\infty} a_k (z-z_0)^{k\alpha}. \quad (4.21)$$

From (3.44), the above expression implies that

$$a_k = \frac{1}{(2\pi)^{\alpha} i^{\alpha}} \cdot \frac{1}{\Gamma(1+\alpha)} \oint_C \frac{f(z)}{(z-z_0)^{(k+1)\alpha}} (dz)^{\alpha}, \quad (4.22)$$

where $C: r^{\alpha} < |z-z_0|^{\alpha} < R^{\alpha}$.

Setting $C_1: |z-z_0|^{\alpha} = r^{\alpha}$ and $C_2: |z-z_0|^{\alpha} = R^{\alpha}$, from (2.44) we have

$$f(z) = \frac{1}{(2\pi)^{\alpha} i^{\alpha}} \cdot \frac{1}{\Gamma(1+\alpha)} \oint_{C_2} \frac{f(z)}{(z-z_0)^{(k+1)\alpha}} (dz)^{\alpha} - \frac{1}{(2\pi)^{\alpha} i^{\alpha}} \cdot \frac{1}{\Gamma(1+\alpha)} \oint_{C_1} \frac{f(z)}{(z-z_0)^{(k+1)\alpha}} (dz)^{\alpha}$$

Successively, it follows from the above that

$$f(z) = \sum_{k=-\infty}^{\infty} a_k (z-z_0)^{k\alpha}, \quad (4.23)$$

where

$$a_k = \frac{1}{(2\pi)^{\alpha} i^{\alpha}} \cdot \frac{1}{\Gamma(1+\alpha)} \oint_C \frac{f(z)}{(z-z_0)^{(k+1)\alpha}} (dz)^{\alpha}, \quad (4.24)$$

for $C: r^{\alpha} \leq |z-z_0|^{\alpha} \leq R^{\alpha}$.

**Theorem 11**

If $f(z)$ has local fractional analytic at every other point inside a contour $C$ with center at $z_0$, then $f(z)$ has a local fractional Laurent series about $z=z_0$ so that

$$f(z) = \sum_{i=-\infty}^{\infty} a_k (z-z_0)^{k\alpha}, 0 < \alpha \leq 1, \quad (4.25)$$

where for $C: r^{\alpha} < |z-z_0|^{\alpha} < R^{\alpha}$ we have

$$a_k = \frac{1}{(2\pi)^\alpha i^\alpha} \cdot \frac{1}{\Gamma(1+\alpha)} \oint_C \frac{f(z)}{(z-z_0)^{(k+1)\alpha}} (dz)^\alpha . \qquad (4.26)$$

*Proof.* Setting $C_1 : |z-z_0|^\alpha = r^\alpha$ and $C_2 : |z-z_0|^\alpha = R^\alpha$, from (2.44) we have that

$$f(z) = \frac{1}{(2\pi)^\alpha i^\alpha} \cdot \frac{1}{\Gamma(1+\alpha)} \oint_{C_2} \frac{f(\xi)}{(\xi-z_0)^\alpha} (d\xi)^\alpha - \frac{1}{(2\pi)^\alpha i^\alpha} \cdot \frac{1}{\Gamma(1+\alpha)} \oint_{C_1} \frac{f(\xi)}{(\xi-z_0)^\alpha} (d\xi)^\alpha . \qquad (4.27)$$

Taking the right side of (4.27) into account implies that for $\xi \in C_2$

$$\left| \frac{(\xi-z_0)^\alpha}{(z-z_0)^\alpha} \right| = \frac{|\xi-z_0|^\alpha}{R^\alpha} = q^\alpha < 1 \qquad (4.28)$$

and

$$|f(\xi)| \le M . \qquad (4.29)$$

By using (4.29) it follows from (4.27) that

$$\begin{aligned}&\frac{1}{(2\pi)^\alpha i^\alpha} \cdot \frac{1}{\Gamma(1+\alpha)} \oint_{C_2} \frac{f(\xi)}{(\xi-z_0)^\alpha} (d\xi)^\alpha \\ &= \frac{1}{(2\pi)^\alpha i^\alpha} \cdot \frac{1}{\Gamma(1+\alpha)} \sum_{n=0}^{\infty} \left[ \oint_{C_2} \frac{f(\xi)}{(\xi-z_0)^{(n+1)\alpha}} (d\xi)^\alpha \right] (z-z_0)^{n\alpha}.\end{aligned} \qquad (4.30)$$

From (4.27) we get

$$\begin{aligned}&-\frac{1}{(2\pi)^\alpha i^\alpha} \cdot \frac{1}{\Gamma(1+\alpha)} \oint_{C_1} \frac{f(\xi)}{(\xi-z_0)^\alpha} (d\xi)^\alpha \\ &= \frac{1}{(2\pi)^\alpha i^\alpha} \cdot \frac{1}{\Gamma(1+\alpha)} \sum_{n=0}^{N-1} \left[ \oint_{C_1} \frac{f(\xi)}{(\xi-z_0)^{(-n+1)\alpha}} (d\xi)^\alpha \right] (z-z_0)^{-n\alpha} + R_N\end{aligned} \qquad (4.31)$$

where

$$\lim_{N \to \infty} R_N = \lim_{N \to \infty} \frac{1}{(2\pi)^\alpha i^\alpha} \cdot \frac{1}{\Gamma(1+\alpha)} \sum_{n=N}^{\infty} \left[ \oint_{C_1} \frac{f(\xi)}{(\xi-z_0)^{(-n+1)\alpha}} (d\xi)^\alpha \right] (z-z_0)^{-n\alpha}$$

is reminder.

Since $|f(\xi)| \le M_1$, taking $\left| \frac{\xi-z_0}{z-z_0} \right|^{n\alpha} = q^{n\alpha} < 1$, we have

$$\begin{aligned}|R_N| &= \left| \frac{1}{(2\pi)^\alpha i^\alpha} \cdot \frac{1}{\Gamma(1+\alpha)} \sum_{n=N}^{\infty} \left[ \oint_{C_1} \frac{f(\xi)}{(\xi-z_0)^{(-n+1)\alpha}} (d\xi)^\alpha \right] (z-z_0)^{-n\alpha} \right| \\ &\le \frac{1}{(2\pi)^\alpha} \cdot \frac{1}{\Gamma(1+\alpha)} \sum_{n=N}^{\infty} \left[ \oint_{C_1} \frac{|f(\xi)|}{|(\xi-z_0)^\alpha|} \left| \frac{\xi-z_0}{z-z_0} \right|^{n\alpha} (d\xi)^\alpha \right]\end{aligned}$$

$$\leq \frac{1}{(2\pi)^\alpha} \cdot \frac{1}{\Gamma(1+\alpha)} \sum_{n=N}^{\infty} \left[ \oint_{C_1} \frac{|M_1|}{|(\xi-z_0)^\alpha|} \left|\frac{\xi-z_0}{z-z_0}\right|^{n\alpha} (d\xi)^\alpha \right]$$

$$\leq \frac{1}{(2\pi)^\alpha} \cdot \frac{|M_1|}{\Gamma(1+\alpha)} \sum_{n=N}^{\infty} \left[ \oint_{C_1} \frac{1}{|(\xi-z_0)^\alpha|} \left|\frac{\xi-z_0}{z-z_0}\right|^{n\alpha} (d\xi)^\alpha \right]$$

$$\leq \frac{1}{(2\pi)^\alpha} \cdot \frac{|M_1|}{\Gamma(1+\alpha)} \sum_{n=N}^{\infty} \left[ \oint_{C_1} \frac{1}{|(\xi-z_0)^\alpha|} q^{n\alpha} (d\xi)^\alpha \right]$$

$$\leq \frac{1}{(2\pi)^\alpha} \cdot \frac{|M_1|}{\Gamma(1+\alpha)} \sum_{n=N}^{\infty} \left[ (2\pi)^\alpha q^{n\alpha} \right] \tag{4.32}$$

$$\leq \frac{|M_1|}{\Gamma(1+\alpha)} \frac{q^{n\alpha}}{1-q^\alpha}.$$

Furthermore

$$\lim_{N\to\infty} R_N = 0.$$

Hence

$$-\frac{1}{(2\pi)^\alpha i^\alpha} \cdot \frac{1}{\Gamma(1+\alpha)} \oint_{C_1} \frac{f(\xi)}{(\xi-z_0)^\alpha} (d\xi)^\alpha$$
$$= \frac{1}{(2\pi)^\alpha i^\alpha} \cdot \frac{1}{\Gamma(1+\alpha)} \sum_{n=0}^{\infty} \left[ \oint_{C_1} \frac{f(\xi)}{(\xi-z_0)^{(-n+1)\alpha}} (d\xi)^\alpha \right] (z-z_0)^{-n\alpha}. \tag{4.33}$$

Combing the formulas (4.30) and (4.33), we have the result.

Hence, the proof of the theorem is finished.

# 5 Generalized residue theorems

In this section we start with a local fractional Laurent series and study generalized residue theorems.

**Definition 8**

Suppose that $z_0$ is an isolated singular point of $f(z)$. Then there is a local fractional Laurent series

$$f(z) = \sum_{i=-\infty}^{\infty} a_k (z-z_0)^{k\alpha} \tag{5.1}$$

valid for $|z-z_0|^\alpha \leq R^\alpha$. The coefficient $a_{-1}$ of $(z-z_0)^{-\alpha}$ is called the generalized residue of $f(z)$ at $z=z_0$, and is frequently written as

$$\operatorname*{Re}_{z=z_0} s\, f(z). \tag{5.2}$$

One of the coefficients for the Yang-Taylor series corresponding to

$$\phi(z) = (z-z_0)^{n\alpha} f(z), \qquad (5.3)$$

the coefficient $a_{-1}$ is the residue of $f(z)$ at the pole $z = z_0$. It can be found from the formula

$$\operatorname*{Res}_{z=z_0} f(z) = a_{-1} = \lim_{z \to z_0} \frac{1}{\Gamma(1+n\alpha)} \frac{d^{(n-1)\alpha}}{dz^{(n-1)\alpha}} \left\{ (z-z_0)^{n\alpha} f(z) \right\} \qquad (5.4)$$

where $n$ is the order of the pole.

Setting $f(z) = \sum_{i=-\infty}^{\infty} a_k (z-z_0)^{k\alpha}$, the expression (5.3) yields

$$\begin{aligned}\phi(z) &= (z-z_0)^{n\alpha} \sum_{i=-\infty}^{\infty} a_k (z-z_0)^{k\alpha} \\ &= a_{-n} + a_{-n+1}(z-z_0)^{\alpha} + a_{-1}(z-z_0)^{(n-1)\alpha} + \ldots\end{aligned} \qquad (5.5)$$

We know that this is

$$a_{-1} = \frac{\phi^{(n-1)\alpha}(z_0)}{\Gamma(1+n\alpha)}, \qquad (5.6)$$

which is the coefficient of $(z-z_0)^{(n-1)\alpha}$.

The generalized residue is thus

$$\operatorname*{Res}_{z=z_0} f(z) = a_{-1} = \frac{\phi^{(n-1)\alpha}(z_0)}{\Gamma(1+n\alpha)}, \qquad (5.7)$$

where $\phi(z) = (z-z_0)^{n\alpha} f(z)$.

**Corollary 12**

If $f(z)$ is local fractional analytic within and on the boundary $C$ of a region $\Re$ except at a number of poles $a$ within $\Re$, having a residue $a_{-1}$, then

$$\frac{1}{(2\pi)^{\alpha} i^{\alpha} \Gamma(1+\alpha)} \oint_C f(z)(dz)^{\alpha} = \operatorname*{Res}_{z=z_0} f(z). \qquad (5.8)$$

*Proof.* Taking into account the definitions of local fractional analytic function and the pole we have local fractional Laurent's series

$$f(z) = \sum_{i=-\infty}^{\infty} a_k (z-z_0)^{k\alpha} \qquad (5.9)$$

and therefore

$$f(z) = \cdots + a_{-n}(z-z_0)^{-n\alpha} + \cdots + a_{-1}(z-z_0)^{-\alpha} + a_0 + \cdots + a_n(z-z_0)^{n\alpha} + \cdots. \qquad (5.10)$$

Hence we have the following relation

$$\frac{1}{\Gamma(1+\alpha)} \oint_C f(z)(dz)^{\alpha} = \frac{1}{\Gamma(1+\alpha)} \oint_C \left( \sum_{i=-\infty}^{\infty} a_k (z-z_0)^{k\alpha} \right)(dz)^{\alpha}. \qquad (5.11)$$

furthermore

$$\frac{1}{\Gamma(1+\alpha)}\oint_C f(z)(dz)^\alpha = \frac{1}{\Gamma(1+\alpha)}\oint_C \frac{a_{-1}}{(z-z_0)^\alpha}(dz)^\alpha. \qquad (5.12)$$

From (3.9), it is shown that

$$\frac{1}{(2\pi)^\alpha i^\alpha}\cdot\frac{1}{\Gamma(1+\alpha)}\oint_C f(z)(dz)^\alpha = \frac{1}{(2\pi)^\alpha i^\alpha}\cdot\frac{1}{\Gamma(1+\alpha)}\oint_C \frac{a_{-1}}{(z-z_0)^\alpha}(dz)^\alpha = a_{-1}. \qquad (5.13)$$

Hence we have the formula

$$\frac{1}{\Gamma(1+\alpha)}\oint_C f(z)(dz)^\alpha = (2\pi)^\alpha i^\alpha a_{-1}. \qquad (5.14)$$

Taking into account the definition of generalized residue, we have the result.

This proof of the theorem is completed.

From (5.8), we deduce the following corollary:

**Corollary 13**

If $f(z)$ is local fractional analytic within and on the boundary $C$ of a region $\Re^\alpha$ except at a finite number of poles $z_0, z_1, z_2\ldots$ within $\Re^\alpha$, having residues $a_{-1}, b_{-1}, c_{-1}\ldots$ respectively, then

$$\frac{1}{(2\pi)^\alpha i^\alpha \Gamma(1+\alpha)}\oint_C f(z)(dz)^\alpha = \sum_{i=0}^n \operatorname{Res}_{z=z_k} f(z) = a_{-1} + b_{-1} + c_{-1} + \ldots. \qquad (5.15)$$

It says that the local fractional integral of $f(z)$ is simply $(2\pi)^\alpha i^\alpha$ times the sum of the residues at the singular points enclosed by the contour $C$.

# 6 Applications: Gauss formula of complex function

**Theorem 14**

Suppose that $f(z)$ is local fractional analytic and $\omega$ is any point, then for the circle

$$|z-\omega|^\alpha = |R^\alpha E_\alpha(i^\alpha \theta^\alpha)|$$

we have

$$f(\omega) = \frac{1}{(2\pi)^\alpha}\cdot\frac{1}{\Gamma(1+\alpha)}\int_0^{2\pi} f(\omega + RE(i\theta))(d\theta)^\alpha. \qquad (6.1)$$

*Proof.* By using (3.1) there exists a simple closed contour $C$ and $z_0$ is any point interior to $C$ such that

$$f(\omega) = \frac{1}{(2\pi)^\alpha i^\alpha}\cdot\frac{1}{\Gamma(1+\alpha)}\oint_C \frac{f(z)}{(z-\omega)^\alpha}(dz)^\alpha. \qquad (6.2)$$

When $C$ can been taken to be $\omega^\alpha + R^\alpha E_\alpha(i^\alpha \theta^\alpha)$ for $\theta \in [0, 2\pi]$, substituting the relations

$$(z-\omega)^\alpha = R^\alpha E_\alpha(i^\alpha \theta^\alpha) \qquad (6.3)$$

and

$$(dz)^{\alpha} = i^{\alpha} R^{\alpha} E_{\alpha}\left(i^{\alpha}\theta^{\alpha}\right)(d\theta)^{\alpha}, \quad (6.4)$$

in (6.2) implies that

$$f(\omega) = \frac{1}{(2\pi)^{\alpha} i^{\alpha}} \cdot \frac{1}{\Gamma(1+\alpha)} \oint_{C} \frac{f(\omega + RE(i\theta)) i^{\alpha} R^{\alpha} E_{\alpha}\left(i^{\alpha}\theta^{\alpha}\right)(d\theta)^{\alpha}}{R^{\alpha} E_{\alpha}\left(i^{\alpha}\theta^{\alpha}\right)} \quad (6.5)$$

and some cancelling gives the result.